\documentclass[superscriptaddress,twocolumn,showpacs,prb,amsmath,amssymb]{revtex4}% Physical Review B
 \usepackage{graphicx}% Include figure files \usepackage{bm}% bold math

  \DeclareMathOperator{\Ree}{Re} \DeclareMathOperator{\Imm}{Im} 
\begin{document}

\def\Xint#1{\mathchoice %Principal value of an integral
   {\XXint\displaystyle\textstyle{#1}}%
   {\XXint\textstyle\scriptstyle{#1}}%
   {\XXint\scriptstyle\scriptscriptstyle{#1}}%
   {\XXint\scriptscriptstyle\scriptscriptstyle{#1}}%
   \!\int}
\def\XXint#1#2#3{{\setbox0=\hbox{$#1{#2#3}{\int}$}
     \vcenter{\hbox{$#2#3$}}\kern-.5\wd0}}
\def\ddashint{\Xint=}
\def\dashint{\Xint-}

\title{Divergence of the effective mass near a density  wave instability in a  MOSFET system.}

\author{V. M. Galitski}
\affiliation{Condensed Matter Theory Center, Department of Physics,
University of Maryland, College Park, MD
20742-4111}
\author{V. A. Khodel}
\affiliation{Condensed Matter Theory Center, Department of Physics,
University of Maryland, College Park, MD
20742-4111}
\affiliation{Russian Research Center ``Kurchatov Institute,'' 123182, Moscow, Russia}

\begin{abstract}
We study the renormalization of the Fermi-liquid parameters in the
vicinity of a density wave quantum phase transition, which should
occur in MOSFET systems at low densities. First, using a perturbative
RPA treatment of fluctuations, we calculate the electronic self-energy
and show that the effective mass diverges at the density wave
transition point. Second, we go beyond perturbation theory, making use
of the exact Pitaevski\^{\i} identities. Within this exact analysis,
we also find a divergence of the effective mass, which occurs at
higher densities in the fluctuation region, as compared to the
perturbation theory.  This result signals the break-down of
conventional Fermi-liquid description in the vicinity of the
transition point. The divergence of the effective mass gives rise to a
singular behavior of the electronic compressibility.  We suggest that
the experimentally observed enhancement of the effective mass is a
precursor to a second order thermodynamic phase transition into a
glassy density wave state.

\end{abstract}


\maketitle

\section{Introduction}
Novel experiments in two-dimensional high-mobility electron systems
have provided a great variety of remarkable results which do not fit
into the framework of the conventional theory of fermion
systems.\cite{Review} The most prominent effect is indeed the
metal-insulator crossover observed in such systems, which clearly
contradicts to the scaling theory of localization \cite{Localization}
and a lot of attempts have been made to reveal the underlying nature
of the phenomenon.  Apart from the unexpected metallic regime, there
is a number of other surprising experimental results, such as the
sharply increasing effective mass at low electron densities
\cite{m1,m2,Tsui} in silicon metal-oxide-semiconductor field-effect
transistors (MOSFETs). There are also experiments clearly showing some
unusual behavior in thermodynamic properties such as the local
electronic compressibility, \cite{compress} suggesting the existence
of a second order thermodynamic phase transition in the system, which,
if indeed exists, should be related to the quantum crossover observed
in the transport behavior.

A successful theory describing the peculiar physics of the
high-mobility electron systems must be based on the correct choice of
the quantum ground state of the system. It is clear now that
 traditional paramagnetic Landau Fermi liquid theory in its
perturbative form does not provide an adequate explanation of the
observed physics.  One is forced to assume the existence of a
density-driven quantum phase transition(s) in the two-dimensional
electron system.

The competition between the potential and kinetic energies of a pure
electron system yields the first order phase transition into the
Wigner solid phase at very low densities. This fact was realized a
very long time ago, \cite{Wigner} but the theoretical description of
the liquid-solid phase transition is still missing and only numerical
studies are available. These numerical studies derive extremely high
values of the Wigner transition RPA parameter: $r_{\rm s} \approx 110$
and $r_{\rm s} \approx 37$ in three \cite{WC3D} and two \cite{WC2D}
dimensions correspondingly. There are also numerical works
\cite{CDW1,CDW2,CDW3} which suggest that in between  Wigner crystal
and  usual Landau Fermi-gas there may be an intermediate phase, the
charge-\cite{RevCDW} or spin-density wave \cite{RevSDW} phase, in
which a stable charge (spin) density modulation exists in the ground
state with some wave-vector $Q \lesssim 2p_{\rm F}$: 
\begin{equation}
\label{CDW} n_{\uparrow \downarrow}({\bf r}) = n_0\left[ 1 + \Delta 
\cos\left({\bf Q} {\bf r} + \delta_{\uparrow \downarrow }\right)\right], 
\end{equation} 
where for the charge-density wave: $\delta_{\uparrow} =
\delta_{\downarrow}$ and for the spin-density wave: $\delta_{\uparrow}
- \delta_{\downarrow} = \pi$. Parameter $\Delta$ serves as the order
parameter for the phase transition. Usually, a density wave transition
is associated with the Peierls instability in effectively
one-dimensional systems,\cite{RevCDW,RevSDW} which occurs due to the
strong electron-phonon coupling. Nevertheless, such a state may appear
in higher dimensional electron systems within a jellium-like model, as
was first pointed out in Ref.~[\onlinecite{Overhauser}].  At low
electron densities, there is a region where a density wave state is
energetically more favorable compared both to the uniform density
distribution corresponding to usual Landau Fermi gas and to Wigner
crystal. \cite{CDW1,CDW2,CDW3} Numerical simulations of
\'{S}wierkowski {\em et al.} \cite{CDW1} showed that in double-layer
and multi-layer systems, the transition into the charge-density wave
state occurs at $r_{\rm s} \sim 10 \mbox{---} 20$, {\em i.e.} much
earlier than the Wigner crystal phase transition is expected. The
charge-density wave state in an isotropic system is possible only if
the neutralizing ``jellium'' is polarizable, while the spin-density
wave should occur for any elasticity of the positive background. In a
MOSFET system, the only possible candidate for such a polarizable
background is the charge distributed on the metallic gates, which is
separated from the two-dimensional system of interest by an oxide
dielectric barrier, usually ${\rm Si O_2}$, of some finite thickness
$d$.  Indeed, if the thickness is very large compared to the typical
inter-electron distance, the ``jellium'' on the gates does not ``see''
any changes of the electron density at the length-scale of $1/p_{\rm
  F}$ and any charge-density modulations should become unstable.
However, if the oxide layer is thin enough, there should exist a
critical density at which a phase transition into the charge-density
phase occurs. This corresponds to the experiments of Tsui {\em et
  al.},\cite{Tsui} where a divergence of the effective mass was
reported for a thin-oxide MOSFET.  In the majority of other
experiments, the thickness of the oxide is about $d \sim
10^3\,\mbox{\AA}$, which is much greater than the typical
inter-electron distance, and the transition into the spin-density wave
phase is anticipated. Thus, independently of the geometry of a
two-dimensional structure, a density-wave phase may exist separating
the paramagnetic Fermi liquid and Wigner solid phases.

The subject matter of the present paper is to explore the behavior of
the Landau Fermi-liquid theory parameters as one approaches a density
wave transition.  Our main finding is that the Fermi-liquid theory
apparently breaks down in vicinity of the transition point as the
effective mass diverges, which leads to the instability of the usual
Fermi-step density distribution even at $T=0$. We suggest that the
observed enhancement of the effective mass \cite{m1,m2,Tsui} is a
precursor of a density wave transition in the two-dimensional systems.

Our paper is structured as follows: In Sec.~II, we study the
renormalization of the quasiparticle $Z$-factor and the effective mass
within the self-consistent RPA approximation in the vicinity of the
charge-density wave transition. This method of treating fluctuations
is in the spirit of the classical paper of Doniach and
Engleberg,\cite{DE} who considered the ferromagnetic transition case,
and also similar to Ref.~[\onlinecite{Chubukov}], where the
self-energy function was studied near the antiferromagnetic transition
in connection with the physics of high-$T_{\rm c}$ cuprates.  First,
we derive general expressions for the  self-energy function.
Second, we discuss the structure of the appropriate RPA density-wave
fluctuation propagator.  The actual calculations are done for the
charge density wave case and it is shown that {\em within the RPA
  approach} the quasiparticle $Z$-factor vanishes at criticality ({\em
  c.~f.}, Ref.~[\onlinecite{Chubukov}]) and the effective mass
diverges as $1/Z$ at the critical point.  In Sec.~III, we study the
fluctuation effects going beyond conventional RPA perturbation scheme, by
using some exact formulae from the Landau Fermi-liquid theory. Namely,
on the basis of the Pitaevski\^{\i} identities, we prove that for a
density-wave phase transition (and more generally, for any phase
transition, which breaks the translational invariance of an otherwise
isotropic system) the renormalization of the derivative
$\left({\partial \Sigma / \partial \omega}\right)$ is never singular
and the $Z$-factor formally never vanishes as distinct of the RPA
result. The effective mass does diverge in the fluctuation region near
the phase transition, but not at the criticality itself, again,
opposed to the RPA conclusion.  At this point, the group velocity of
quasiparticles changes sign and the conventional Fermi-liquid theory
breaks down.\cite{KS,Noz} Moreover, at this point some thermodynamic
quantities, such as the electronic compressibility, experience a
singular behavior.  In Sec.~IV, which is the conclusive part, we
summarize our findings and briefly discuss our understanding of
possible physics in the density-wave phase.

\section{RPA-treatment of the density-wave fluctuations}

\subsection{Fermi-liquid theory distilled}

 The effective Fermi-liquid parameters follow from the quasiparticle Green function, defined as usual:\cite{Mahan}
 \begin{equation}
 \label{Gdef}
 G_{\rm c}\left( t_1, {\bf r}_1; t_2, {\bf r}_2 \right) = - i 
 \left\langle {\rm T} \psi \left( t_1, {\bf r}_1\right)  \psi^{\dagger} \left( t_2, {\bf r}_2 \right) \right\rangle,
 \end{equation}
 where $T$ is the time-ordering operator and $\psi$ and $\psi^\dagger$
 are electron field operators. In what follows, we will use subscript
 ``c'' to denote a time-ordered quantity and subscripts ``R'' and
 ``A'' to denote a retarded and advanced quantities correspondingly. The Green
 function (\ref{Gdef}) is in turn determined by the self-energy
 function $\Sigma\left(\varepsilon,{\bf p}\right)$ as follows:
\begin{equation}
\label{G1}
G_{\rm c} \left(\varepsilon,{\bf p}\right) = {1 \over \varepsilon - 
\xi^{(0)}\left({\bf p}\right) - \Sigma_{\rm c}\left(\varepsilon,{\bf p}\right)}.
\end{equation}
In the vicinity of the Fermi-momentum, Eq.~(\ref{G1}) can be re-written as
\begin{equation}
\label{G}
G_{\rm c} \left(\varepsilon,{\bf p}\right) = {Z \over \varepsilon - 
\xi\left({\bf p}\right) \pm i \gamma\left(\varepsilon,{\bf p}\right)},
\end{equation}
where $Z \in \left[ 0\, , \, 1\right]$ is the jump in the
quasiparticle distribution function at the Fermi level, which is
determined by the following identity:
\begin{equation}
\label{Z}
Z^{-1} = 1 - \left[ {\partial \over \partial \varepsilon} \Ree \Sigma_{\rm R} \left(\varepsilon,p_{\rm F}\right) \right] \Biggr|_{\varepsilon = 0}
\end{equation}
and $\xi\left({\bf p}\right) = \left( \left| {\bf p} \right| - p_{\rm
  F} \right) p_{\rm F}/m_*$ is the quasiparticle spectrum, with the
effective mass defined via the following relation
\begin{equation}
\label{m*}
{m_* \over m} = {1 \over Z} \left\{ 1 - \left[ {\partial \over \partial \xi_{\bf p}} \Ree 
\Sigma_{\rm R} \left( 0,p \right)\right]\Biggr|_{p=p_{\rm F}}   \right\}^{-1}.
\end{equation}
The parameter $\gamma = Z \Imm \Sigma_{\rm c}\left(\varepsilon,p_{\rm
  F}\right)$ determines the quasiparticle life-time and should be
small compared to the energy of a quasiparticle, for the latter to be
well-defined. Under this assumption, we can write the retarded Green
function in the following form, which shall be used in further actual
calculations:
\begin{equation}
\label{G2}
G_{\rm R}  \left(\varepsilon,{\bf p}\right)  = \mbox{\Large \rm P}\, {Z \over \varepsilon - 
\xi\left({\bf p}\right)} - i \pi Z \delta \left[ \varepsilon -  \xi\left( {\bf p} \right) \right],
\end{equation}
where  symbol $\mbox{ \rm P}$ stands for the principal value of the corresponding integral.                  

Within the perturbative treatment of fluctuations, only the first
order diagram (containing one fluctuation propagator) for the
self-energy is retained.  The corresponding analytically continued
retarded self-energy function reads (for $\varepsilon >0$):\cite{AGD}
\begin{eqnarray}
\label{Sig_gen}
\nonumber
&&\!\!\!\!\!\!\!\!\!\!\!\! \Sigma_{\rm R} \left( \varepsilon, {\bf p} \right) = 
-{1 \over \left( 2 \pi \right)^3 }
\int d^2 {\bf q}
\int\limits_{-\infty}^{+\infty} d \omega \\ \nonumber
&& \times \Biggl\{ {\rm Im\,} G_{\rm R} \left( \omega + \varepsilon, {\bf p} - {\bf q} \right)
D_{\rm R}\left(-\omega , {\bf q} \right) \tanh \left[ {\omega + \varepsilon
  \over 2 T} \right] \\&&
\,\,\,\,\,\,  +\, G_{\rm R} \left( \omega + \varepsilon, {\bf p} - {\bf q} \right)
 {\rm Im\,} D_{\rm R}\left(\omega, {\bf q} \right) \coth \left[ {\omega  \over 2 T} \right] \Biggr\}.
\end{eqnarray}
The interaction $D\left(\omega,{\bf q}\right)$ changes the properties of the system and renormalizes
the Fermi-liquid theory parameters. In the case under discussion, the density
wave fluctuation propagator will be used as the appropriate effective interaction term.

From now on, we will consider zero temperature case only, so that the tangent and cotangent factors
in Eq.~(\ref{Sig_gen}) reduce to theta-functions. Moreover, using Eqs.~(\ref{G2}) and (\ref{Sig_gen}),
we can evaluate the integral over the directions of ${\bf q}$ exactly (provided that the interaction is isotropic) 
and obtain the following expression for the real part of the self-energy, which is quite involved:
\begin{equation}
\label{ReS}
\Ree \Sigma_{\rm R} \left( \varepsilon, {\bf p} \right) =
 \Sigma_1 \left( {\bf p},\varepsilon \right) + \Sigma_2 \left( {\bf p},\varepsilon \right),
\end{equation}
where
\begin{eqnarray}
\label{S1}
\Sigma_1 \left( \varepsilon, {\bf p} \right) = {Z \nu \over \pi} \dashint\limits_{-\infty}^{\infty} 
\! d \omega\, {\rm sgn \,} \left( \omega + \varepsilon \right) \int\limits_{q_1\left(\Omega\right)}^{q_1\left(\Omega\right)}
dq \nonumber \\ 
\,\,\,\,\,\, \times {q    \Ree D_{\rm R} \left( \omega,q \right)
\over \sqrt{ \left[ q^2 - q_1^2\left( \Omega \right) \right]  \left[  q_2^2\left( \Omega \right) - q^2 \right] }}
\end{eqnarray}
and
\noindent \begin{eqnarray}
\label{S2} \nonumber
\Sigma_2 \left(\varepsilon, {\bf p} \right) = -{Z \nu \over \pi} \dashint\limits_{-\infty}^{\infty} 
\! d \omega\,  {\rm sgn \,}   \omega \left\{ \int\limits_0^{q_1\left(\Omega\right)} dq\, q
+ \int\limits_{q_2\left(\Omega\right)}^\infty dq \right\} \\
\!\!\!\! \times 
 {   \Imm D_{\rm R} \left( \omega,q \right)
\over \sqrt{ \left[ q^2 - q_1^2\left( \Omega \right) \right]  \left[   q^2 - q_2^2\left( \Omega \right) \right] }},
\end{eqnarray}
where
\begin{equation}
\label{q12}
q_{1,\, 2} \left( \Omega \right) = \sqrt{2} p_{\rm F} \left[ \left( 1 + {\Omega \over 2 E_F} \right)  \mp 
\left( 1 + {\Omega \over E_F} \right)^{1/2} \right]^{1/2}
\end{equation}
and $\Omega = \left(  \omega + \varepsilon - \xi_{\bf p} \right)$, with $\nu$ being the density of states at the Fermi-line
and $E_F$ the Fermi energy.

Using Eqs.~(\ref{ReS}), (\ref{S1}), and (\ref{S2}), one can prove the following useful identity:
\begin{equation}
\label{iden}
\left[ {\partial \over \partial \varepsilon} + {\partial \over \partial \xi_{\bf p}} \right] 
\Sigma_{\rm R} \left(\varepsilon, {\bf p} \right)   \Biggr|_{ \stackrel {\displaystyle  p=p_{\rm F}} {\displaystyle \varepsilon=0}} \equiv
{2 Z \nu \over \pi} \int\limits_0^{2 p_{\rm F}} {dq\, D \left( 0, q\right)\over \sqrt{ (2 p_{\rm F} )^2 - q^2}}. 
\end{equation}
Let us note, that the right-hand side of Eq.~(\ref{iden}) is very large, provided that the effective propagator is 
singular in the static limit.
 
\subsection{Charge-density wave fluctuation propagator.} 
 
Now let us consider a thin-oxide MOSFET in which the two-dimensional
electron system is separated from the metallic gates by a barrier of
width $d$, which we suppose to be not too large compared to the
typical inter-electron distance. In this case, the charge on the
metallic gates serves as an elastic (not rigid) neutralizing
background, which can compensate the energy cost produced by
inhomogeneities in the electron density distribution in the
two-dimensional electron system. At sufficiently low electron
densities, one anticipates a transition into the charge-density wave
state (\ref{CDW}) which should be energetically more favorable than
the homogeneous distribution. Below the critical density $n_{\rm c}$
the initial Galilean invariance of the system is broken. However,
above the transition the system is still isotropic and only
fluctuating density waves can exist with no preferable direction. The
appropriate {\em static } fluctuation propagator, quite generally has
the form:
\begin{equation}
\label{D1}
D\left(\omega=0,q\right) \propto -{1 \over \xi^{-2} + \left(q - Q\right)^2 },
\end{equation}
where $\xi$ is the charge-density wave coherence length, which
diverges at the transition and $Q \sim 2 p_{\rm F}$.  The static part
of the propagator (\ref{D1}) has a typical Lorentzian form (see, {\em
  e~g.}, Ref.~[\onlinecite{Dyugaev}]) and, as we noted, should be
still isotropic above the transition as the Galilean invariance is
preserved until the transition point: $n=n_{\rm c}$.  The transition itself is thermodynamic
in nature and is all about the sign of the difference between the two
energies of the homogeneous distribution and distribution (\ref{CDW}).
The RPA philosophy assumes that the dynamic part of the fluctuation
propagator, which is not crucial for the transition itself, can be
obtained by summing up a series of diagrams containing the
RPA polarizability bubbles, which generate the required frequency
dependence.  Taking into account these considerations, we write the
RPA charge-density wave fluctuation propagator in the following form:
\begin{equation}
\label{D}
D\left(\omega,q\right) = -{\lambda \over \nu_0}\, {1 \over \alpha + \left[1 - q/Q \right]^2 + 
i P\left( \omega,q\right)},
\end{equation}
where $\alpha \sim \left(n -n_{\rm c} \right)/n_{\rm c}$ is the Ginzburg-Landau-like coefficient, which changes sign at the transition point,
$\nu_0$ is the bare density of states at the Fermi-line and  the imaginary part $iP$ has the form
$$
 P\left( \omega,q\right) = {\Lambda \over \nu_0} \Imm \Pi \left(\omega,q\right)
 $$
where $\Pi\left( \omega,q\right)$ is the polarizability bubble and  $\lambda$ and $\Lambda$ are some dimensionless constants, which may depend on 
$r_{\rm s}$ only. The polarizability is defined as usual (here, we use  Matsubara notations):
\begin{equation}
\label{Pmat}
{ \mbox{\Large $\pi$}}\left( \omega_m,{\bf q}\right) = 2 \sum_{\varepsilon_n} \int {d^2{\bf k} \over \left( 2 \pi \right)^2 }
{\cal G}\left( \varepsilon_n,{\bf k}\right){\cal G}\left(\varepsilon_n+\omega_m,{\bf k}+{\bf q}\right).
\end{equation}
For a two-dimensional electron gas at $T=0$, this quantity was
calculated by Stern \cite{Stern}. We do not need its exact expression.
Let us only emphasize, that the case $Q \sim 2 p_{\rm F}$ may be
different from the case $Q \equiv 2 p_{\rm F}$, since the
polarizability is a non-analytic function of  frequency at $q=2
p_{\rm F}$ and the mixture of the two non-analyticities makes the
problem technically more cumbersome. (for Stern's
polarizability:\cite{Stern} $\Imm \Pi\left (\omega, 2 p_{\rm F}
\right) \sim \sqrt{\left| \omega \right|} {\rm sgn\,} \omega$).  Quite
generally, $Q < p_{\rm F}$,\cite{CDW1,CDW2} in which case we can
expand the propagator at low frequencies to obtain:
$$
P (\omega)   \approx \nu  Z^2\left( \omega \over  \omega_0\right),\,\, \mbox{as }\omega\to 0,
$$
where $\omega_0$ is a constant, which does not depend on the
closeness to the critical point, $Z$ is the quasiparticle $Z$-factor,
and $\nu=m_*/\left(2 \pi \right)$ is the density of states, both to be
found.

In the case of a spin-density wave transition, which may occur in a
MOSFET with a large oxide barrier, the propagator should have the form
similar to (\ref{D1}) but with a non-trivial spin structure:
\begin{equation}
\label{DS}
D^{\rm SDW}_{\alpha\beta} \left(\omega,q\right) \propto  {\sigma_{\alpha \gamma} \sigma_{\gamma \beta} \over \xi^{-2} + 
\left(q - Q\right)^2 + i \tilde{P} \left( \omega,q\right)},
\end{equation}
In what follows, we will do explicit calculations for the case of
the charge-density wave only. Generalization to the spin-density wave
case is straightforward and the qualitative results are identical.

\subsection{Renormalized $Z$-factor and effective mass}

 Now, we are at the position to calculate the renormalization of the Fermi liquid parameters due to charge-density fluctuations 
 within the RPA-approach.  Identity (\ref{iden}) allows us to calculate only one derivative, 
 the other will be automatically extracted from Eq.~(\ref{iden}). 
Let us focus on the $\varepsilon$-derivative, which determines the renormalization  of the quasiparticle $Z$-factor. From  Eq.~(\ref{S1}), we have
\begin{eqnarray}
\label{S12}
{\partial \over \partial \varepsilon} \Sigma_1 \left( \varepsilon, {\bf p} \right) = {Z \nu \over \pi} \int\limits_{-\infty}^{\infty} 
\!\!\! d \omega\, {\rm sgn \,}\omega \int\limits_{q_1\left(\omega\right)}^{q_1\left(\omega\right)} &&
\!\!\!\!\!\! {   dq\, q
\over \sqrt{ \left( q^2 - q_1^2\right)  \left(  q_2^2 - q^2 \right) }} \nonumber \\
&&\!\!\!\!\!\!\!\!\!\!\!\!\!\!\!\!\!\!\!\!\!  \times {\partial \over \partial \varepsilon}  \Ree D_{\rm R} \left( \omega -\varepsilon ,q \right),
\end{eqnarray}
where functions $q_{1,\,2}\left( \omega \right)$ are defined in
Eq.~(\ref{q12}).  Let us consider the region in the immediate vicinity
of the transition only so that $\alpha \ll \left( 2 p_{\rm F} - Q
\right) / 2 p_{\rm F}$. Then, the integrals in Eq.~(\ref{S1}) and
(\ref{S2}) are determined by low frequencies $\omega \sim \alpha/ Z^2
\to 0$. Using the following properties of any physical propagator
$$
\Imm D_{\rm R}\left( \omega,{\bf q}\right) = - \Imm D_{\rm R}\left( -\omega,{\bf q}\right)
$$
and
$$
\Ree D_{\rm R}\left( \omega,{\bf q}\right) =  \Ree D_{\rm R}\left( -\omega,{\bf q}\right),
$$
one can see that expression (\ref{S1}) is identical to the
right-hand side of  Eq.~(\ref{iden}).  Eq.~(\ref{S2}) is exactly
zero as the principal value of the corresponding integral.  In the
leading order, we obtain the following results:
\begin{equation}
\label{Res_eps}
{\partial \over \partial \varepsilon}  \Ree \Sigma \left( \varepsilon, p_{\rm F} \right)  \Biggl|_{\varepsilon =0 }
= - {2 \lambda \over \sqrt{1 - \left(Q/ 2 p_{\rm F} \right)^2}} {Z \over \sqrt{\alpha}}
\end{equation}
and 
\begin{equation}
\label{Res_xi}
{\partial \over \partial \xi_{\bf p}}  \Ree \Sigma \left( 0, p \right)   \Biggl|_{p = p_{\rm F} } = \overline{\overline{\it o}} \left( \alpha \right) \to 0,
\end{equation}
where we remind that $\alpha = \left(n - n_{\rm c} \right) /n_{\rm c}$
is a small deviation from the charge-density wave transition point.

Let us mention that besides the singular fluctuation effects we
studied so far there are other contributions coming from the $q \to 0$
channel. Such corrections have been considered previously by many
authors (see {\em e.g.}, Ref.~[\onlinecite{TLQ}]) and it was found
that they do increase the effective mass by a factor of two or so
(Ting {\em et al.}\cite{TLQ} predict $m_*/m\approx 2.0$ for $r_{\rm s}
= 5$) but do not lead to any divergent behavior.  Let us denote the
corresponding renormalized effective mass as $\tilde{m}$ and the
quasiparticle $Z$-factor as $\tilde{Z}$.  In these notations we obtain
the following self-consistent equation for the quasiparticle
$Z$-factor within the RPA approach [we consider here the limit
$\alpha \ll \left(2 p_{\rm F}/Q - 1 \right)^2$]:
 \begin{equation}
 \label{ZEq}
 {2 \lambda \over \sqrt{1- \left(Q / 2 p_{\rm F}\right)^2}} {Z^2(\alpha) \over \sqrt{\alpha}} + {Z(\alpha) \over \tilde{Z}} - 1 =0.
 \end{equation}
 This quadratic equation can be easily solved exactly, but we are
 mostly interested in the behavior in the closest vicinity of the
 transition, which leads to ({\em c.~f.},
 Ref.~[\onlinecite{Chubukov}]):
 \begin{equation}
 \label{Zas}
 Z\left( \alpha \to 0\right) = {1 \over \sqrt{2 \lambda}} \left[ 1 - \left( {Q \over 2 p_{\rm F}} \right)^2 \right]^{1 \over 4} \, \alpha^{1/4} \propto
 \left( n - n_{\rm c} \right)^{1/4}.
 \end{equation}
 Thus, the effective mass diverges as $1/Z$, and we have
\begin{equation}
 \label{mas}
 m_* = {\tilde{m}  \over Z(\alpha)} \propto {1 \over \left(n - n_{\rm c} \right)^{1/4}},\,\,\, Q<2 p_{\rm F}.
 \end{equation}
 We see that within the RPA-treatment of charge-density wave
 fluctuations, the effective mass diverges at criticality, while the
 $Z$-factor vanishes.
 
 Let us mention that the case $Q=2 p_{\rm F}$ is somewhat
 pathological, since the pole of the propagator coincides with the
 point of the non-analyticity of the polarizability function and also
 with the point where the inverse square-root term appearing in the
 two-dimensional integrals (\ref{S1}) and (\ref{S2}) diverges. To
 obtain the correct numerical factors and to avoid unphysical
 divergences, one should use the exact form of the propagator in the
 hole range of the $(\omega,{\bf q})$-space.  If we are not interested
 in exact numbers, we can easily estimate the scaling law
 of the $Z$-factor near the transition for the $Q=2 p_{\rm F}$ case
 within the RPA approximation:
 \begin{equation}
 \label{Zas2p}
 Z_{Q=2p_{\rm F}} \left( \alpha \right) \sim \lambda^{1/2} \alpha^{3/8}
  \end{equation}
 Thus, the effective mass diverges as
 \begin{equation}
 \label{mas2p}
 m_* \propto {1 \over \left( n -n_{\rm c}\right)^{3/8}},\,\,\,\, Q=2p_{\rm F}
 \end{equation}
 Let us emphasize that  results (\ref{Zas}----\ref{mas2p}) take into account only one diagram, which
 alone yields very divergent results. 
 
 \section{Non-perturbative analysis.}
 
 Although, the RPA treatment is widely-accepted and often leads to
 reasonable results, we believe that in some cases there are serious
 doubts concerning its reliability. Unlike in the electron-phonon
 problem, where the Migdal theorem holds and circumvents the necessity
 to study complex vertex corrections, some other problems (such as,
 {\em e.~g.}: paramagnon coupling near a magnetic phase
 transition,\cite{Chubukov,DE} charge density wave transition, spin
 density wave transition, {\em etc}.) require correct account for the
 vertex corrections, which are not small in the vicinity of the
 critical point and the Migdal theorem is apparently violated.  The
 exact account for the vertex corrections seems an insurmountable
 problem at the moment and the usual practice is to cover the
 underlying difficulties by saying that the domain of applicability of
 the RPA-like treatment may be extended and that the perturbative
 approach should qualitatively explain the key physics of the
 transition. In the present section, we prove that such an arguing is
 quite dangerous and may lead to some qualitatively incorrect
 conclusions.  Fortunately, the density-wave transition case may
 uncover the underlying problem, as one can obtain some very important
 results based on very general grounds.  As we show below, in the
 vicinity of any phase-transition, which breaks the Galilean
 invariance of an initially isotropic system, some important effects
 are hidden in higher order diagrams. The main qualitative result we
 are going to derive in the present section is that the effective mass
 diverges before the phase transition occurs and this divergence leads
 to the break-down of the Fermi-liquid theory. Apart from the
 perturbative result, predicting $Z=0$ at the criticality, we show
 that the $Z$-factor formally always remains finite within the
 non-perturbative treatment.
 
 Paradoxally, we will use the essence of the Fermi-liquid theory
 construction to justify its break-down. Let us introduce such a standard
 element as the irreducible four-vertex function \cite{AGD} $
 \Gamma\left(p_1;p_2|p_1+k; p_2+k\right)$, where we introduce
 notations $p_i = \left(\varepsilon_i, {\bf p}_i\right)$ for brevity, with
 ${\bf p_1}$ and ${\bf p_ 2}$ being the momenta of  incoming and
 outcoming particles correspondingly. In the limit $k = \left(\omega,
 {\bf k} \right) \to 0$ the four-vertex has a singular structure and we,
 following Ref.~[\onlinecite{AGD}] introduce the following standard
 function:
 \begin{equation}
 \label{Gw}
 \Gamma^{\omega} \left(p_1;p_2 \right) = \lim\limits_{\stackrel{\omega \to 0} {|\bf k|/\omega \to 0}} \Gamma\left(p_1;p_2| p_1+k; p_2+k\right).
 \end{equation}

 We also decompose the product of two Green functions involved in the
 calculation of the four-vertex $\Gamma$ into a singular and regular
 parts:\cite{AGD}
 \begin{equation}
 \label{GG}
 G(p) G(p+k) = {2 \pi i Z^2 \over v} {{\bf v k} \over \omega - {\bf vk} } \delta(\varepsilon) \delta\left(\left|{\bf p}\right| - p_{\rm F}\right) + g_{\rm reg}(p),
 \end{equation}
 where $g_{\rm reg}(p)$ is a regular function in the limit ${\bf k},\,\omega \to 0$.
 One of the central  identities of the formal Landau Fermi liquid theory construction reads\cite{Pitaevski}
 \begin{equation}
 \label{id1}
 {\partial\, G^{-1}\left( p \right)  \over \partial \varepsilon} = {1 \over Z} = 1 + {1 \over 2} \int 
  \Gamma^{\omega} \left(p,p' \right) g_{\rm reg}(p')\, {d^{\left(2+1\right)} p' \over i\left(2  \pi \right)^3}.
 \end{equation}
 The other identities, which we refer to as the Pitaevski\^{\i}
 identities,\cite{Pitaevski} can be obtained from Eq.~(\ref{id1}) by
 considering symmetries of the initial Hamiltonian (in our case
 isotropic electron system with Coulomb interactions between
 electrons).  The Galilean invariance, which is broken in a density
 wave state, but is preserved above the phase transition yields the
 second Pitaevski\^{\i} identity which reads:\cite{AGD,Pitaevski}
  \begin{equation}
 \label{id2}
 {\bf p} {\partial\, G^{-1}\left( p \right)  \over \partial \varepsilon} = {\bf p}  +{1 \over 2} \int 
  \Gamma^{\omega} \left(p ; p' \right) {\bf p'} g_{\rm reg}(p')\, {d^3 p' \over i \left(2  \pi \right)^3}
 \end{equation}
 From Eqs.~(\ref{id1}) and (\ref{id2}), the following relations follow  immediately:
\begin{equation}
\label{paradox1}
 {\partial   \Sigma  \left(\varepsilon,p_{\rm F}\right)  \over \partial \varepsilon}   \Biggr|_{\varepsilon = 0} =
  {1 \over 2} \int 
  \Gamma^{\omega} \left(p ; p' \right) g_{\rm reg}(p')\, {d^3 p' \over i \left(2  \pi \right)^3}
  \end{equation}
 and
 \begin{equation}
\label{paradox2}
 {\partial   \Sigma  \left(\varepsilon,p_{\rm F}\right)  \over \partial \varepsilon}   \Biggr|_{\varepsilon = 0} =
   {1 \over 2} \int 
  \Gamma^{\omega} \left(p ; p' \right) {\left( {\bf p} {\bf p'} \right) \over p_{\rm F}^2} g_{\rm reg}(p')\, {d^3 p' \over i \left(2  \pi \right)^3}.
  \end{equation}
  In the case of a phase transition, preserving the Galilean
  invariance of the system, the four-vertex function should be peaked
  at $\left| {\bf p} - {\bf p'}\right| = 0$ and the relations
  (\ref{paradox1}) and (\ref{paradox2}) become identical in the
  leading order, since $\left( {\bf p p'} \right) = p_{\rm F}^2$. In
  the case of a second order phase transition, which does break the
  translational symmetry (of which a density-wave phase transition is
  an excellent example indeed) we anticipate the four-vertex to
  diverge at the transition point at some finite wave-vector $Q$. {\em
    I.~e.}, in the fluctuation region, the major contribution should
  comes from the domain $\left| {\bf p} - {\bf p'}\right| = Q$. Therefore,
  relations (\ref{paradox1}) and (\ref{paradox2}) lead to an
  obvious contradiction, which is especially pronounced in the case $Q
  = 2 p_{\rm F}$, when $\left( {\bf p p'} \right)/ p_{\rm F}^2 = -1$.
  The only possible resolution of the paradox is to accept that
\begin{equation}
\label{dSde=0}
{\partial   \Sigma  \left(\varepsilon,p_{\rm F}\right)  \over \partial \varepsilon}   \Biggr|_{\varepsilon = 0} = 0
\end{equation}
in the leading order in the closeness to the phase transition. We see
that this exact result is in some sense opposite to the one obtained
within the perturbative treatment [see Eq.~(\ref{Zas})], which proves
that the RPA-approach contains a serious deficiency. Obviously, the
singularity can not just disappear from the theory, since the general
Landau relation\cite{LL} ensures that at least one of the derivatives
$\left( d \Sigma / d \varepsilon \right)$ and $\left( d \Sigma / dp \right)$ is singular [{\em
  c.~f.,} Eq.~(\ref{iden})]. The relation reads
\begin{eqnarray}
\label{Lrel}
\left[ {\partial \over \partial \varepsilon} + {\partial \over \partial \xi_{\bf p}} \right] 
\Sigma \left(\varepsilon, {\bf p} \right)   \Biggr|_{ \stackrel {\displaystyle  p=p_{\rm F}} {\displaystyle \varepsilon=0}} =
{m \over Z} \int && \!\!\!\!\!\!\!  {d^2{\bf p}  \over \left(2 \pi p_{\rm F}\right)^2} f\left({\bf p},{\bf p}'\right) \nonumber \\
&& \!\!\!\!\!\!\!\!   \times\, {\partial n({\bf p}') \over \partial {\bf p}'}  {\bf p} ,
\end{eqnarray}
where we introduced the full scattering amplitude (Landau function), which is the static four-vertex function:
$$
 f\left({\bf p,p}'\right)  =   \Gamma^{\omega}\left( 0, {\bf p}; 0, {\bf p'} \right). 
 $$
 In the perturbative approach, this singularity arose in the
 $\varepsilon$-derivative of the self-energy. We see, that within the
 very general non-perturbative analysis the singularity is
 ``transferred'' into the $p$-derivative, which automatically means
 that the quasiparticle $Z$-factor never vanishes and the only
 quantity which acquires singular contributions and eventually
 diverges is the effective mass. The equation determining effective
 mass renormalization can be derived from Eq.~(\ref{id1}) and the
 gauge invariance of the system, and in the most convenient form can be
 written as:
  \begin{equation}
  \label{Eq-m}
  {{\bf p}  \over m} = {\partial \xi_{\bf p}[n] \over \partial {\bf p} }  -  \int  f\left({\bf p -p'}\right)  {\bf p} {\partial n ({\bf p}') \over \partial {\bf p}'}
  {d ^2 {\bf p}' \over \left( 2 \pi \right)^2},
  \end{equation}
  where $n({\bf p})$ is the electron distribution, which does not have
  to be the Fermi-distribution function, in general.  $\xi_{\bf p}[n]
  $ is the electron spectrum, which in general is a complex functional
  of the distribution function.\cite{LL,KS,Noz} Ideally, the electron
  distribution function should be self-consistently determined from
  the Landau energy functional, as its absolute minimum and must be subject to
  the stability constraint $ {\partial \xi_{\bf p}[n] / \partial p > 0
  }$ ({\em i.~e.}, $m_*>0$).  However, as long as the group velocity
  of the quasiparticles (the inverse effective mass) is positive, the
  familiar Fermi-step solution is the correct distribution function at
  $T=0$.
  
  The scattering amplitude $f$ is a singular quantity in the vicinity
  of the transition and quite generally can be written as:
  \begin{equation}
  \label{stvertex}
 f\left({\bf p -p'}\right) = -{\lambda \over \nu_0} 
  \left\{ \alpha + \left[ {\left| {\bf p} - {\bf p}'\right| - Q \over Q} \right]^2 \right\}^{-1},
  \end{equation}
  where again $\nu_0 = m/\left( 2 \pi \right)$ is the ``bare'' density
  of states at the Fermi line, $\alpha \propto \left( n - n_{\rm
    c}\right)$ is a small parameter, which changes sign at the
  transition, $\lambda$ is a dimensionless constant, which may depend
  on $r_{\rm s}$ only, and $\left(2 \pi/ Q \right)$ is the spatial
  period of the density wave modulations.  Eq.~(\ref{Eq-m}) leads to
  the following formula for the effective mass:
  \begin{equation}
  \label{LFL}
  {1 \over m} = {1 \over m_*} -{2 \lambda \over m \pi } \int\limits_0^{\pi} 
  \left[ \alpha + \left( {2 p_{\rm F} \over Q} \sin{\chi \over 2} - 1\right)^2 \right]^{-1}\!\! \cos\chi d\chi.
  \end{equation}
  After elementary integration we obtain the following result for $\sqrt{2} p_{\rm F} < Q < 2 p_{\rm F}$:
  \begin{equation}
  \label{Q<2p}
  m_* = m \left[ 1 - 2 \lambda {\left| Q^2/\left(2 p_{\rm F}^2\right) - 1 \right| \over \sqrt{\left(2 p_{\rm F} / Q \right)^2 -1}} {1 \over \bf \sqrt{\bf \alpha}}
  \right]^{-1}
  \end{equation} 
  and for $Q = 2 p_{\rm F}$:
    \begin{equation}
  \label{Q=2p}
  m_* = m \left[ 1 - 2 \lambda \alpha^{-3/4} \right]^{-1}.
  \end{equation} 
  We see that in the vicinity of the transition $\alpha \to 0$, the
  effective mass diverges, independently on the value of $\lambda$,
  and formally changes sign after this point in the fluctuation
  region. This immediately signals that the Fermi-step distribution
  function is not stable\cite{KS,Noz} and conventional Fermi liquid
  theory breaks down.
  
  Another quantity, which can be easily estimated from the general
  considerations of the Landau Fermi liquid theory is the electronic
  compressibility, which can be expressed as\cite{LL}
$$
\left({\delta \mu \over  \delta n}\right) \propto {1 \over m_*} \left( 1 +  \nu  \overline{f\left(\chi\right)}\right),
$$
where $\nu = m_* / \left( 2 \pi \right)$ is the density of states
and $\overline{f\left(\chi\right)}$ is a singular quantity in the
vicinity of the phase transition [see Eq.~(\ref{stvertex})].
Therefore, we conclude that the electronic compressibility $\left|
\partial \mu / \partial n\right|$ must show a divergent behavior in
the vicinity of the phase transition (as $m_* \to \infty$).  Clearly, other
thermodynamic quantities such as specific heat, susceptibility
($g$-factor), {\em etc.} may also acquire a singular behavior. A
detailed analysis of thermodynamics in the vicinity of the charge-
and spin-density wave transitions will be reported elsewhere.

 \section{Conclusion}
 Summarizing, we suggest that the metal-insulator transition may be
 actually a signature of a thermodynamic phase transition in a dilute
 two-dimensional electron system. Namely, the charge- or spin-density
 wave transition.  Which of the two phases prevails is a tentative
 question and the answer strongly depends on the geometry of the
 system and can be determined only on the basis of an accurate numerical
 analysis comparing the energies of the two states.
 Experimentally, the spin-density wave case may be distinguished from
 the charge-density wave one by measuring the electron $g$-factor
 which should be singular in the former scenario and unremarkable in
 the latter.  Our expectation is that in thin-oxide
 MOSFET's,\cite{Tsui} the charge-density wave should be more
 energetically favorable, while the spin-density wave should take over
 if the metallic gates are located  far from the two-dimensional electron
 gas.  In both scenarios, we proved that the density-wave fluctuations
 lead to the divergence of the electron effective mass and to a
 singular behavior of the electron compressibility. Both effects have
 been observed experimentally. \cite{compress,m1,m2,Tsui} The density
 corresponding to $m_* =\infty$ is the point, where the Fermi-liquid
 description in its conventional form apparently breaks down. In the
 immediate vicinity of the density wave transition, the problem
 directly maps on the model of so-called fermion condensation
 introduced earlier by Khodel and Shaginyan\cite{KS} and considered
 later by Nozi{\`e}res\cite{Noz} and also by Volovik.\cite{Volovik}
 The issue of such a non-Fermi-liquid state is far from its closure,
 but the main consequence is that the Fermi-distribution function can
 not be a stable solution of the Landau energy functional if it leads
 to a negative group velocity of the quasiparticles.  Possible model
 distributions were suggested,\cite{KS,Noz} all characterized by a
 very high density of states at the chemical potential $\nu(\mu)$
 (with the singularity smeared out only by temperature), which may
 correspond to the sharp increase of the conductivity.
 
 Below the phase transition, {\em i.~e.}, in a density wave state,
 disorder should play the key role leading to the destruction of the
 long-range order and to the pinning of  density waves (according to the
 general theorem due to A. I. Larkin \cite{pinning}). The glassy
 nature of the density wave state should reveal itself via a noisy
 behavior of linear response quantities (such as conductivity), which
 may show some real-time glassy behavior. The physics here should be
 quite similar to the physics of superconducting vortices in the
 presence of disorder.\cite{RevCDW,RevSDW,Vort} Let us mention
 experiments of Jaroszy{\'n}ski {\em et al.}\cite{Glass1} and of
 Bogdanovich and Popovi{\'c},\cite{Glass2} in which the low-frequency
 resistance noise was measured clearly showing the glassy freezing of
 the electronic system.
 
 Finally, let us emphasize an important technical discovery concerning
 the reliability of a perturbative RPA-like treatment of strong
 fluctuations and the importance of vertex corrections.  The problem
 is quite general and is related, in particular, to the physics of
 fluctuation phenomena near the antiferromagnetic phase transition in
 high-$T_{\rm c}$ cuprates. \cite{Chubukov} We have shown that the
 vertex correction, which are usually excluded from
 consideration,\cite{DE,Chubukov} yield qualitatively important
 changes and ensure that the quasiparticle $Z$-factor does not vanish.
 Our current understanding, based on studying a model example allowing
 parquet technique treatment (to be published elsewhere), is that the
 effect of the vertex corrections is mostly the renormalization
 (suppression) of the imaginary part of the initial RPA-like
 propagator.  This result seems quite reasonable, since the imaginary
 part of the propagator is related to the life-time of fluctuations,
 which should become infinite at the transition point.

\acknowledgments{The authors are grateful to Victor Yakovenko for very useful discussions.
VMG thanks Euyheon Hwang for  valuable comments. VMG is supported by  the US-ONR, LPS, and DARPA.
VAK is supported by the NSF Grant
PHY-0140316 and by the McDonnell Center for Space Sciences, 
 by the Grant 00-15-96590 from the Russian Foundation for Basic
Research, and Grant NS-1885.2003.2 from the Russian Ministry of
Industry and Science.}

\bibliography{cdw}

\end{document}